\documentstyle[aps,epsf]{revtex}

\newcommand{\ket}[1]{\left | \, #1 \right \rangle}
\begin{document}
\title{On the analytical convergence of the QPA procedure}
\author{Chiara Macchiavello}
\address{Dipartimento di Fisica ``A. Volta'' and I.N.F.M., 
Via Bassi 6, 27100 Pavia, Italy}
\maketitle
\begin{abstract}
We present an analytical proof 
of the convergence of the ``quantum privacy amplification'' 
procedure proposed by D. Deutsch {\it et al.} [Phys. Rev. Lett. {\bf 77}, 
2818 (1996)]. The proof specifies the range of states which can be purified 
by this method.
\end{abstract}

Purification schemes of two-particle entangled states have been recently
proposed \cite{B,QPA}. 
The purpose of such schemes is to distill a subset of states with enhanced
purity from a larger set of non pure entangled states of two particles.
They have proved essential to perform various tasks 
in quantum information theory, such as teleportation of quantum states
over noisy channels \cite{B}, secure quantum cryptography in the presence
of noise \cite{QPA}, quantum error correction \cite{BPRA} and distributed
quantum computation \cite{dqc}.
They have also been related to some fundamental problems in quantum mechanics,
such as the separability of quantum states and the nature of entanglement
\cite{h3}.
The most efficient scheme known so far 
for a wide range of the initial states 
is the so called ``quantum privacy amplification'' (QPA). It was originally
designed for cryptographic purposes \cite{QPA}.
In this paper we analyse in detail the QPA algorithm and  present an 
analytical proof of its convergence. This part was missing
in the original paper, where some conclusions were based on numerical analysis.


An elementary step of the QPA protocol is described by the following map
\begin{eqnarray}
  A&=&(a^2+b^2)/p\\
  B&=&2cd/p\\
  C&=&(c^2+d^2)/p\\
  D&=&2ab/p\;,
\end{eqnarray}
where $p=(a+b)^2+(c+d)^2$, while
$\{a,b,c,d\}$ and $\{A,B,C,D\}$ are the diagonal elements of the 
density operator describing the state of a ``noisy'' EPR pair
in the Bell basis representation
$\{\ket{\phi^+},\ket{\psi^-},\ket{\psi^+},\ket{\phi^-}\}$:
\begin{eqnarray}
{\ket{\phi^\pm}  =\frac {1}{\sqrt{2}} (\ket{00}\pm\ket{11})}
\atop
{\ket{\psi^\pm} =\frac {1}{\sqrt{2}} (\ket{01}\pm\ket{10})}\;.
\label{bell_basis}
\end{eqnarray}
The two states $\{\ket{0},\ket{1}\}$ form 
the basis of the two-dimensional systems belonging to the EPR pair. 
The small letters $a,b,c,d$ correspond to the density
operator of the pair before a QPA step, 
while the capital letters $A,B,C,D$ correspond to
the surviving state at the output.
It is understood that normalisation of the density 
operator requires $a+b+c+d=1$, and therefore the map involves only three
independent parameters.

In the following we want to prove analytically that the QPA map converges to 
the value $\{1,0,0,0\}$ (corresponding to the pure state $\ket{\phi^+}$) 
for any initial value $a > 0.5$. In other words, if the initial fraction of
the $\ket{\phi^+}$ component of the density operator at the beginning of the
procedure is larger than 0.5, then the iteration of the QPA algorithm
will asymptotically lead to the final state $\{1,0,0,0\}$.


The proof is based on showing the following assertions:

\begin{itemize}

\item[$i$)] 
 There exists a monotonic function in the region ${\cal R}=\{
a\in (0.5,1],b,c,d\in [0,0.5); a+b+c+d=1\}$, i.e.
it increases under iterations of the QPA map.

\item[$ii)$] 

The extremal value of this monotonic function 
in the above mentioned region corresponds to the fixed point 
of the QPA map in this region, which is $\{1,0,0,0\}$.
                                              

\end{itemize}

The above properties of the map lead to the conclusion
that the fixed point $\{1,0,0,0\}$ is also
an attractor (and the only one) in the region of interest, thus proving 
the convergence of the map for any initial value $a > 0.5$. 

The most natural candidate function one could 
think of is the first component $a$
in the diagonal of the density operator, which asymptotically (for a large 
number of iterations of the procedure) reaches its maximum value 1.
This is the case for example in the first proposed purification scheme of
Ref. \cite{B}, where it was easily shown that the first component always 
increases at each iteration, which immediately proved the convergence of the
method. In the present QPA algorithm this is not the case, as $a$ is not always
smaller than the corresponding $A$.

The function we will consider in this paper has the quadratic form
\begin{eqnarray}
f(a,b)=(2a-1)(1-2b)\;.
\end{eqnarray}
Let us point out in passing that such function depends only on two of the
three independent parameters involved in the map. 
In order to prove the monotonicity of funtion $f$ we will first prove that when
the parameters of the input density operator 
$\{a,b,c,d\}$ belong to the region ${\cal R}$ 
it increases in one iteration of the map, i.e. 
\begin{eqnarray}
f(A,B)>f(a,b)\;
\label{mon}
\end{eqnarray}
and then show that if the initial diagonal values of the density operator 
belong to the region of interest ${\cal R}$, they stay in this region at 
all subsequent iterations of the map.

After some straightforward algebra, condition (\ref{mon}) can be more 
conveniently expressed only in terms of parameters $c$ and $d$ as 
\begin{eqnarray}
g(c,d)&=&2(c+d)^4-4(c+d)^3+4(c+d)^2\nonumber\\
&-&(c+d)-(c^2+d^2)<0\;.
\label{g}
\end{eqnarray}
Notice that also this condition involves only
two of the three independent parameters which characterise the map.
In order to prove Eq. (\ref{g}) we just have 
to prove that function $g$ is always negative in the region 
$\{c,d\in [0,0.5), c+d<0.5\}$.

We can easily see that the function is negative on the boundaries of the 
region of interest, namely on the axis $c=0$ and $d=0$ with $0<c,d<0.5$ and
on the line $c+d=0.5$ (apart from the points $c=d=0$ and $c=d=0.25$ 
where $g=0$).
Moreover, the function is decreasing when we depart from the boundaries towards
the inner part of the interested region. 
We can also show that $g$ has only one extremal point in the 
region, which is a minimum.
This can be seen by introducing the variables $x=c-d$ and $y=c+d$:
the vanishing conditions for the partial derivatives 
${{\partial g}\over {\partial x}}$ and ${{\partial g}\over {\partial y}}$ 
lead to the solution $y=0$ (i.e. $c=d$) and the following equation for $y$
\begin{eqnarray}
8 y^3-12 y^2+7y-1=0\;.
\end{eqnarray}
We can see analytically that the above equation has only one solution,
which lies between 0 and 0.5: the exact value has been found numerically 
to be $y_0=0.205122$. 
As the function $g$ is continuous, the above features guarantee the negativity
of $g$ in the region of interest.
We have thus proved that $f(A,B)>f(a,b)$ for $a > 0.5$ in one iteration of 
the QPA map. 

Moreover, for one iteration of the map we have
\begin{eqnarray}
1-2A=\frac{(2a-1)(2b-1)}{p}\;.
\label{a1/2}
\end{eqnarray}
The above equation shows that $A > 0.5$ if $a > 0.5$,
i.e. if the initial value of $a$ is larger than $0.5$ then it will be always 
larger than $0.5$ at all subsequent iterations of the map and will never cross
the $a=0.5$ boundary line. This means that the evolutions of parameters
$a,b,c$ and $d$ under the QPA map will never leave the region ${\cal R}$.
Therefore, the function $f$ is also always increasing for all iterations
of the map if $a$ is initially larger than 0.5. 
In this way we have proved point $i$).

Regarding point $ii$), it is easy to see that the function 
$f(a,b)$ takes its maximum value for
$f(1,0)=1$ in the region of interest $\{a\in (0.5,1],b\in [0,0.5), 
a+b\leq 1\}$: $f(1,0)$ corresponds to the maximum value along the boundaries
of the region and no local extremal points are present inside. Therefore, by
iterating the QPA map, the function $f$ is bound to reach its maximum value
and the surviving states of the pairs asymptotically approach the pure state 
$\ket{\phi^+}$.
This allows to conclude that whenever the initial value of $a$ is bigger
than $0.5$, the density operator of the pairs is driven
to the fixed point of the map $\{1,0,0,0\}$.
Such point is therefore a global attractor for any initial value $a>0.5$.

\begin{figure}[thb]
\vskip .3truecm\begin{center}
\epsfxsize=.3\hsize\leavevmode\epsffile{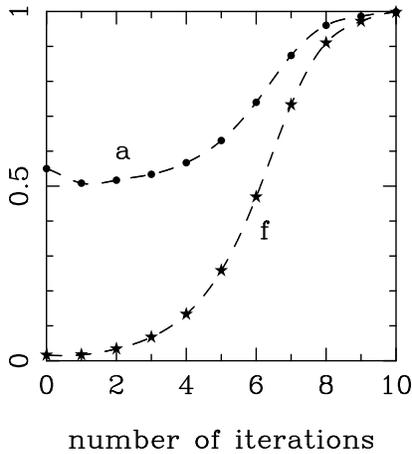}
\end{center}
\caption{Plot of the first component $a$ in the diagonal of the density 
operators (circles) and function $f$ (stars) as functions of the number
of iterations of the QPA map for the initial state $\{0.57,0.41,0.01,0.01\}$.
The zero-th iteration corresponds to the initial values. Please note that
$a$ decreases in the first step of the iteration.}
\label{f:fig1}\end{figure}

As an example, in Fig \ref{f:fig1} we plot the behaviour of $f$
as a function of the number of iterations
for the initial state $\{0.57,0.41,0.01,0.01\}$, in contrast to 
the non monotonic behaviour of the first component $a$ of the density operator.

Let us now analyse what happens when the diagonal elements of the
initial density operator do not belong to the region ${\cal R}$.  From
Eq. (\ref{a1/2}) we can see that after the first iteration of the map
we get $A >0.5$ for any initial value $b>0.5$.  Thus, for 
an initial value of $b$ bigger than 0.5 after one
iteration the parameters $a,b,c$ and $d$ belong to the region ${\cal
R}$ and the map is therefore converging to the state $\{1,0,0,0\}$.
Notice also that the QPA map is symmetric
under the following exchange of parameters $a\leftrightarrow c$,
$b\leftrightarrow d$.  This implies that the proof we have presented
here is valid also for any initial value $c, d>0.5$ and in this
case the map converges to the state $\{0,0,1,0\}$.
We can then conclude that
this kind of map leads to a perfect state purification whenever 
one of the initial coefficients $a,b,c$ or $d$ is larger than 0.5. The 
final pure state is either $\ket{\phi^+}$ if $a$ or $b$ are initially larger 
than 0.5, or $\ket{\psi^+}$ if $c$ or $d$ are initially larger 
than 0.5.

When the initial diagonal components of the density operator are all smaller
than 0.5 it is not possible to purify the state by means of the present
QPA map. Actually, we can see from Eq. (\ref{a1/2}) that
$A$ cannot be greater than 0.5 if both $a$ and $b$ are smaller.
Moreover, for one iteration of the map we have
\begin{eqnarray}
1-2B=\frac{2(c^2+d^2)-2(c+d)+1}{p}\;,
\end{eqnarray}
and this expression is always positive for $c,d<0.5$,
meaning that also $B$ is smaller than 0.5.
Because of the symmetry of the map under exchange 
$a\leftrightarrow c$ and $b\leftrightarrow d$, the same conclusions hold 
for $C$ and $D$. Therefore, when we initially have $a,b,c,d<0.5$, 
then $A,B,C,D<0.5$
and none of the diagonal components of the density operator will ever cross
the 0.5 boundary line at any iteration of the
map. The procedure in this case is not successful.

At the moment the physical meaning of the function $f$ is still unclear. 
Notice that other monotonic functions can be obtained from $f$ by 
any monotony preserving transformation, but still we could not 
give any obvious physical explanation for the convergence of the map.
Let us then leave this as a challenge to our colleagues in the field.

We are grateful to David Deutsch, Artur Ekert and Pave\l 
$\;$ Horodecki for very useful discussions.


\end{document}